\documentclass[pra,twocolumn,amsmath,amssymb]{revtex4-2}
\usepackage{bm}

\usepackage{epsfig}
\usepackage{xcolor}

\begin{document}

\title{Second Born electrons, born again seamen}  
\author{A.\  R.\  P.\  Rau\footnote{ravirau@lsu.edu} \\
Department of Physics and Astronomy, 
Louisiana State University, 
Baton Rouge, Louisiana 70803}

\begin{abstract}

The multiple puns in the title play on a curiosity, that the rescue of a person overboard at sea and the dominance of the second Born term in forward charge transfer in atomic collisions share common elements of physics. Essentials and commonality in the two are explained. 

\end{abstract}


\maketitle
    
\section{Introduction}

$\bullet$  What is the most efficient maneuver when a person falls overboard at sea?

$\bullet$  In charge transfer in the forward direction during ion-atom collisions, the second Born approximation dominates over the first Born at high velocities.

Remarkably, these two items from entirely different areas, one from the classical world of seamen and naval battleships and the other from microscopic quantum mechanics of atoms and electrons, share at heart the same physics. Remarkably as well, a prescient physicist anticipated the second nearly a hundred years ago before the advent of quantum mechanics and well before the naval maneuver developed independently decades later.   

\section{Rescue at sea of person overboard}

When a person falls overboard at sea, time is of the essence in the rescue. For a large ship moving at high speed, shutting off and reversing the engines, ``slamming on the brakes," to come to a stop, then turn around and return, takes a very long time, many ship lengths. More efficient procedures have been developed, varying according to weather and other conditions \cite{ref1}, our concern in this paper with one termed the ``Williamson-Butakov" turn. It seems to have originated from a Russian naval officer before the 1903 Japan-Russia war to turn around battleships and have their guns point in the opposite direction and, independently by a US naval officer during WW II to rescue sailors who have fallen overboard. It came to my attention in an article by the writer John McPhee in The Atlantic Monthly \cite{ref2}, also available as chapter 2 in a book \cite{ref3}. The prescription is to maintain speed but turn hard right (or left) and, when the ship is headed $60^\circ$ from the original path, to turn the steering wheel hard in the opposite direction. This twin turn maneuver results in a circular path that puts the ship back downstream headed exactly in the opposite direction as shown in Fig. 1, all the while at speed and thus in short time. 

Clearly, a repeat at that point will return the ship to the exact location and heading as when the sailor went overboard although this may not be required for the purpose at hand, a straight line path after the first turn sufficing for that return and rescue. Geometrically, Fig. 1 may be viewed in terms of two identical circles of diameter $d$ which is the distance between initial and final positions of the ship indicated by the arrowheads where the circles are tangent to that horizontal direction. The circles are side-by-side to touch (``osculate") on the common tangent which is the vertical bisector of that horizontal separation.

The path of the ship in the Williamson-Butakov turn shown in Fig. 1 divides the total circumference $\pi d$ travelled into two segments, a quarter on the first circle clockwise and then transitioning to the three-quarter circumference of the second circle anti-clockwise. The hard reversal of the steering wheel takes place after the ship has swung through $60^\circ$ (at that point, it is a vertical distance $d/4$ below the horizontal) so as to make a smooth, symmetric transition between the two circular segments, covering $30^\circ$ more on the first circle and then $30^\circ$ on the second. That leaves finally $240^\circ$ on the second to complete the turn. (A second Williamson-Butakov maneuver would complete the very symmetric figure-eight closed circuit over the entirety of the two osculating circles to return to the exact initial location and velocity.) The time taken for a single turn, $\pi d/v$ for a ship speed of $v$, is to be contrasted with a uniform deceleration $a$ along the horizontal to a stop so as to turn around. For typical values of $v$ and $a$ of a large cargo vessel or battleship, the distance $v^2/2a$ travelled and time $v/a$ taken would both be much larger than those for the Williamson-Butakov turn. It is interesting to contrast decelerations of cars and large powered ships, the former reaching approximately the value of $g$, the acceleration due to gravity, while large ships are, remarkably(!), about a thousandth of that. 

\begin{figure}
\scalebox{2.0}{\includegraphics[width=1.6in]{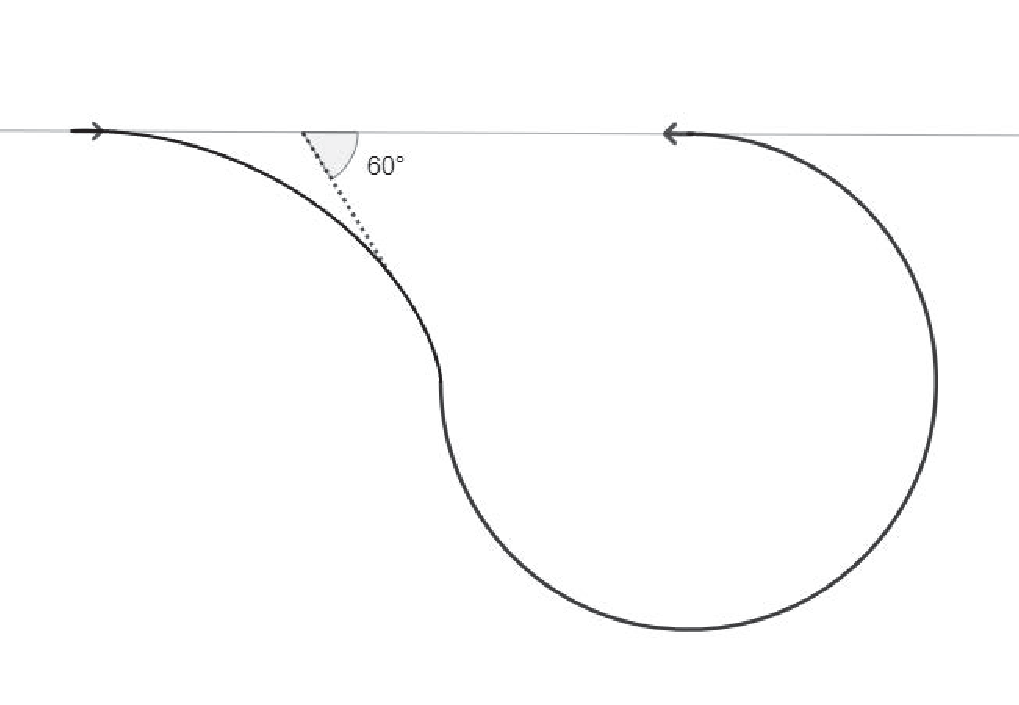}}
\caption{The Williamson-Butakov turn for efficiently turning around a ship at sea. A hard turn at full speed, followed by a hard turn in the opposite direction upon reaching a heading $60^\circ$ to the original puts the ship back downstream facing the opposite direction to its original.}
\end{figure}

\section{Charge transfer collisions and Born approximations}

When an ion is incident on an atom and captures one of its bound electrons, that is referred to as charge transfer. At the simplest, a proton so capturing the electron in a hydrogen atom is depicted as p + H $\rightarrow$ H + p and represents the process when identical nuclei are involved. Capture may be more general, may involve different nuclei, may be accompanied by a change in the bound state or not, may be with or without accompanying radiation, and into either forward or backward direction.  All of these are of interest with applications in aeronomy, astronomy, and plasma physics. There is a large literature on this topic, and an extensive review article \cite{ref4} dealing with various interesting aspects (among them also a pun attributed in \cite{ref4} to Paul Berman who referred to the second Born as ``Born again" approximation) but our concern here is with one, namely the capture in the forward direction at high initial velocity $v$ (while still remaining non-relativistic) of the projectile. 

The first quantum-mechanical treatment of this forward capture, done in the first Born approximation, was by Brinkman and Kramers \cite{ref5} in 1930 and predicted the cross section to drop off as $v^{-12}$. This was in conflict with an early treatment in 1927 through classical mechanics by Thomas \cite{ref6} that had the cross section decrease as $v^{-11}$. Apparently, Bohr and others found this disagreement astonishing, especially given that the cross section for Rutherford scattering of charged particles by one another coincided in quantum and classical treatments, and perturbation theory should be expected to apply. Subtle interference phenomena were suspected and it was not until 1955 that the resolution came from Drisko \cite{ref7} who reasoned that the double scattering involved in the Thomas picture needed as counterpart the second Born term in the perturbation series. Upon evaluation, that second Born term indeed declined as in the classical Thomas result. Thus, with its weaker fall-off in $v$, that term and behavior should prevail at high velocity.

\begin{figure}
\vspace{-1.3in}
\scalebox{2.0}{\includegraphics[width=1.6in]{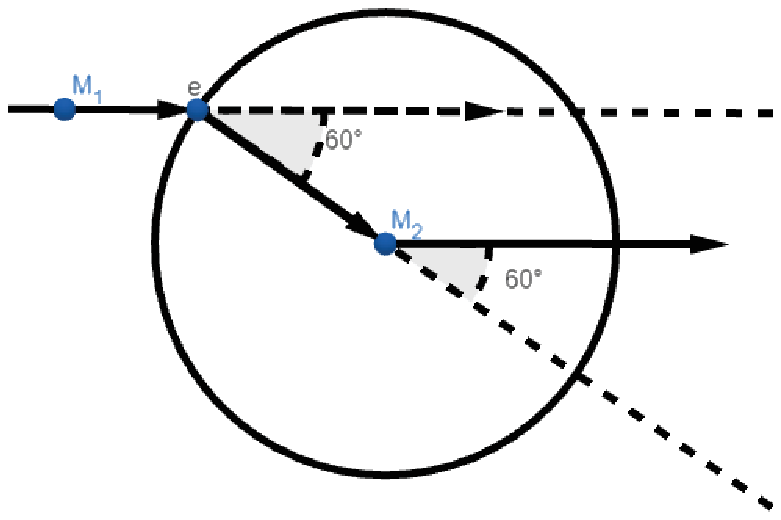}}
\vspace{-1.2in}
\caption{Charge capture by an incident heavy ion $M_1$ of an electron initially bound to an ion $M_2$ requires two $60^\circ$ collisions as initially pointed out by Thomas. The electron is driven by the first to collide with its parent ion so as to end travelling with the same velocity for easy capture by the incident projectile. Kinematics shown underly the dominance of the second Born over the first term in the perturbation series describing such forward ``charge transfer."}
\end{figure}
 
The Born approximation goes back a long way, to Born who also developed alongside the probability interpretation of the wave function that is central to quantum physics. It is essentially a perturbative approach and dominates to this day in quantum scattering in all areas of physics. The related work of Born and Oppenheimer dominates molecular physics, exploiting the very different velocities and therefore time scales of electronic and nuclear motions in a molecule to separate the treatment of them. 

The Born approximation for scattering, usually regarded as valid for high velocities, rests on the effect of scattering being weak. When one body scatters off another with high $v$, the time of interaction is small, justifying the expectation of a weak effect so that perturbation theory should apply. The incident plane wave of definite linear momentum acquires only a small change so that, to a first approximation, the matrix element governing the scattering can be evaluated by integrating the interaction potential between two plane waves describing both initial and final states. This is relatively easy, indeed reduces in essence to a Fourier transform of that interaction and is therefore employed widely in physics. It is often termed a ``form factor". Calculating this first term in a perturbation series of the interaction which thereby enters only once in the transition matrix element constitutes the first Born approximation and is used throughout all areas of physics. To proceed further to the second term in the perturbation series would require considering the modified plane wave in calculating the matrix element, so that the interaction occurs twice. Such a ``distorted wave" calculation is enormously more complicated because that distortion is a sum over all intermediate states, just as in any second order term in perturbation theory. The number of intermediate states being usually infinite, second order calculations are much more difficult. Given the much greater complexity of calculating the second order term both analytically and numerically, most calculations stop with the first Born and that itself referred to as the Born approximation. For other aspects of the Born approximation, its early history, and use by pioneers such as Bohr and Bethe, see Sec. I of a review \cite{ref8}.

In a perturbation series, when a second order term dominates the first order one, an immediate question that arises is about even higher terms and the convergence of the perturbation series. General considerations such as this, of the Born series and its convergence, are still not definitively settled as indeed so for many a perturbation series we use in physics. However, as discussed by Dettman and Leibfried \cite{ref9} and in great detail in \cite{ref4}, there are good reasons to believe that in this case of forward capture, the second Born does give the correct result, a bound placed on the sum of higher terms also pointing this way. The correct fall-off for forward capture is $v^{-11}$ in agreement with Thomas's classical result. Further, when one considers different initial bound states from which capture happens into varied final states, in the limiting case when both are high Rydberg states of the electron, the second Born term gives even the exact result including pre-factors \cite{ref10}, as one would expect when the electron is almost ``free" and quasi-classical schemes become exact for very weakly bound Coulomb states. Thus, it is well established that in forward capture, the second Born prevails over the first at high velocities \cite{ref4}.     

\section{Common elements of the two problems, role of conservation laws}

It was Thomas's intuition that forward capture requires a double collision, the incident ion projectile striking the electron in the target atom to have it then collide with its parent ion, both times being deflected at $60^\circ$ to line up with the projectile to be picked up by it. (This also has a gravitational counterpart in astrophysics.) As shown in Fig. 2, such a kinematics is necessary given the very different mass of ion and electron. See \cite{ref11} and a more detailed figure for the $60^\circ$ angle: Fig. 3 in \cite{ref4}. A single collision from behind would simply shoot the light electron forward with very high velocity and no pick-up by the projectile ion would be possible. The same $60^\circ$ occurs in both figures and in both rescue at sea and forward charge exchange.

In quantum physics, there are no trajectories but the reason these figures and our classical intuition based on them remain valid is because the conclusions depend only on geometry and conservation laws of energy and momentum which are equally valid in a classical and a quantum world. Indeed, even in our first acquaintance in a first-year physics class of collisions between two equal billiard balls or of a cannon ball hitting a golf ball, diagrams made to aid our imagination are only stand-ins for applying energy and momentum conservation in elastic collisions. And, an undergraduate exercise applied to Fig. 2 as two elastic collisions in which a projectile of mass $M_1$ and a light mass $m$ end finally with the same velocity $\vec{V_1}$ while mass $M_2$ initially at rest ends up with $\vec{V_2}$, $m$ moving in between with $\vec{v_e}$ (all these being two-dimensional vectors), will provide through energy-momentum conservation six equations for those six unknowns in the three vectors. See Fig.1 of \cite{ref12}. The $60^\circ$ angle of $\vec{v_e}$ can be easily verified and, of course, the magnitudes of $V_1$ and $v_e$ will be nearly equal, and both nearly equal to initial $v$ while $M_2$ will move almost vertically down with a very small speed. 

In the quantum physics of scattering, the intermediate $\vec{v_e}$ of the electron, trajectory, and $60^\circ$ angle are not accessible at infinity and thus not measurable quantities. But the slight deflection of projectile ion (and captured electron, both being the direction of $\vec{V_1}$) from initial incident direction in Fig. 2, which also follows from the above kinematics and equals $(m/M_1) \sin 60^\circ $, is accessible and this $\approx 0.47$ mrad has been measured experimentally \cite{ref13,ref14}. An observed peak in the scattering cross-section, termed the Thomas peak, at that angle confirms in detail the Thomas mechanism and that $60^\circ$ angle. Other details such as an asymmetric cusp have also been investigated \cite{ref15} but are not relevant to the discussion here. 

Turning to the Williamson-Butakov turn which also involves two turns and the $ 60^\circ $ angle, there is no invoking of conservation laws of physics to account for that particular number but only an argument of symmetry, along with the response time of large ships to hard turns of the steering wheel. That seems to be why the two turns are made as shown in Fig. 1 when the tangent to the circle reaches $60^\circ$. Note also that different considerations apply from those for powered ships when a sailboat is involved in a man overboard maneuver because of the importance of maintaining a heading into the wind at all stages of the turn in order to avoid capsizing \cite{ref16}. See a so-called Q-turn in \cite{ref1}.   

The fundamental laws of conservation, geometry, and symmetry provide the explanation for the close connection (in the poet William Blake's words:``To see the world in a grain of sand, And all eternity in an hour") between two otherwise entirely different phenomena discussed here while validating Thomas's early intuition that was realized later in terms of the dominance of the second term over the first in the Born series. 

\section{Acknowledgment}

I thank Khristian Tallent for drawing the figures. 

I dedicate this essay to the memory of my post-doctoral father, Larry Spruch (1923-2006) of New York University, who taught me many things in and out of physics. Somewhat related to the title of this essay is his being shocked that I did not know how to swim. He gave me swimming lessons during a stay at the Aspen Center one summer fifty years ago.

\end{document}